# Energy and Network Aware Workload Management for Geographically Distributed Data Centers

Ninad Hogade, *Student Member, IEEE,* Sudeep Pasricha, *Senior Member, IEEE,* and Howard Jay Siegel, *Fellow, IEEE*

**Abstract** — Cloud service providers are distributing data centers geographically to minimize energy costs through intelligent workload distribution. With increasing data volumes in emerging cloud workloads, it is critical to factor in the network costs for transferring workloads across data centers. For geo-distributed data centers, many researchers have been exploring strategies for energy cost minimization and intelligent inter-data-center workload distribution separately. However, prior work does not comprehensively and simultaneously consider data center energy costs, data transfer costs, and data center queueing delay. In this paper, we propose a novel game theory-based workload management framework that takes a holistic approach to the cloud operating cost minimization problem by making intelligent scheduling decisions aware of data transfer costs and the data center queueing delay. Our framework performs intelligent workload management that considers heterogeneity in data center compute capability, cooling power, interference effects from task co-location in servers, time-of-use electricity pricing, renewable energy, net metering, peak demand pricing distribution, and network pricing. Our simulations show that the proposed game-theoretic technique can minimize the cloud operating cost more effectively than existing approaches.

**Index Terms** — geo-distributed data centers, workload management, peak shaving, net metering, network cost, game theory, Nash equilibrium

———————————— ◆ ————————————

## 1 INTRODUCTION

In recent years, the ever-increasing use of smartphones, wearable devices, portable computers, Internet-of-Things (IoT) devices, etc., has fueled the use of cloud computing. Data-intensive applications in the domains of artificial intelligence, distributed manufacturing systems, smart and connected energy, and autonomous vehicles are beginning to leverage cloud computing extensively [1]. To support these applications, cloud service providers are increasingly deploying new data centers that can bolster the capabilities of their cloud computing offerings. While centralized data centers were common in the past, more recently there has been a trend towards deploying data centers across geographically diverse locations [2], [3]. Distributing data centers geographically across the globe leads to many benefits. It brings them closer to customers offering better performance (e.g., latency) and lower network costs. It also provides better resilience to unpredictable failures (e.g., environmental hazards) due to the redundancy that distributed data centers enable.

Another strong motivation to geographically distribute data centers is to reduce electricity costs by exploiting time-of-use (TOU) electricity pricing [4]. The cost of electricity varies based on the time of day and follows the TOU electricity pricing model [5]. Electricity prices are higher when the total electrical grid demand is high and falls during periods when electrical grid demand is low [6]. For non-residential or commercial customers, utilities also charge an additional flat-rate (peak demand) charge based on the highest (peak) power consumed at any instant during a billing period, e.g., month [7]. In some cases, such peak demand charges can be higher than the energy use portion of the electric bill. Moving workloads among geo-distributed data centers is one effective approach to reduce electricity expenditures. This approach allows cloud service providers to allocate workloads to data centers with lower energy costs. This not only reduces operating costs for cloud service providers, but also can reduce cloud computing costs for customers.

Distributing workloads geographically entails an overhead: network costs for transferring workloads and their data between data centers. Many workloads hosted in geographically distributed data centers need to transfer data among data centers for data replication, collection, and synchronization. Such data movement significantly increases the cloud networking costs. Even though the price of Internet data transfers continues to decline by approximately 30% per year [8], inter-data center traffic is exploding [9]. Moreover, operating energy-efficient data centers at high capacity can sometimes overwhelm the intra-data center servers and networking equipment, causing a delay in processing of the incoming workload. We call this the data center queueing delay, and this reduces cloud provider performance and revenues.

Data centers today are energy intensive, and are estimated to account for around 1% of worldwide electricity use. The total energy used by the world's data centers has doubled over the past decade and some studies claim that it will triple or even quadruple within the next decade [10]. We define cloud operating costs as the dollar energy costs for all geo-

• N. Hogade is with the Dept. of Electrical and Computer Engineering, Colorado State University, Fort Collins, CO 80523. E-mail: ninad.hogade@colostate.edu.
• S. Pasricha and H.J. Siegel are with the Dept. of Electrical and Computer Engineering, and the Dept. of Computer Science, Colorado State University, Fort Collins, CO 80523. E-mail: {sudeep, hj}@colostate.edu.

distributed data centers plus the dollar inter-data center data transfer costs Minimizing such costs is important as the annual electricity expenditure for powering data centers is growing rapidly: China's data centers alone are on track to use more energy than all of Australia by 2023 [11]. These energy expenses annually can sometimes even surpass the costs of purchasing all of the data center equipment.

It should be noted that geo-distributed data centers have significant heterogeneity in operating costs and performance, because of aspects including: (a) inter-data center network costs, that are affected by the amount of data transferred among sites; (b) queueing delay caused by the intra-data center network congestion within the data centers executing at high capacity; (c) variable TOU pricing, as data centers are often located in different times zones; (d) the use of on-site green/renewable energy sources, e.g., solar and wind, to various extents across sites; (e) the availability (or absence) of net metering, which is a mechanism that gives renewable energy customers credit on their utility bills for the excess clean energy they sell back to the grid [12]; (f) variable peak demand pricing from utility providers across sites; (g) the availability of diverse resources within a data center, such as cooling infrastructure, and heterogeneity across compute nodes (e.g., from the perspective of power and performance characteristics); and (h) co-location interference, a phenomenon that occurs when multiple cores within the same multicore processor are executing tasks simultaneously and compete for shared resources, e.g., last-level caches or DRAM. It is important to factor in this heterogeneity while making decisions about workload management in geo-distributed data centers.

The goal of this work is to design and evaluate a geographical workload distribution solution for geo-distributed data centers that will minimize the cloud operating cost for executing incoming workloads considering all of the aspects mentioned above. This work is applicable to environments where execution information about the workloads is readily available or can be predicted by some form of a workload prediction technique, e.g., [13], [14]. Examples of such environments that exist in the industry include commercial companies (DigitalGlobe, Google), military computing installations (Department of Defense), and government labs (National Center for Atmospheric Research). More specifically, we propose a novel game theory-based workload management framework that distributes workload among data centers over time, while meeting their performance requirements. The novel contributions of our work can be summarized as follows:

- we formulate the cloud workload distribution problem as a non-cooperative game and design a new Nash equilibrium based intelligent game-theoretic workload distribution framework to minimize the cloud operating cost;
- our framework simultaneously considers energy and network cost minimization, while satisfying workload performance goals;
- our decisions leverage detailed models for a comprehensive set of characteristics that impact cloud operating costs and workload performance, including data center compute and cooling power, co-location performance interference, TOU electricity pricing, renewable energy, net metering, peak demand pricing distribution, data center queueing delay, and the costs involved with inter-data center data transfers.

We organize the rest of the paper as follows. In Section 2, we review relevant prior work. We characterize our system model in Section 3. Sections 4 and 5 describe our specific problem and the framework we propose to solve it. The simulation environment is discussed in Section 6. Lastly, we analyze and evaluate the results of our approach in Section 7.

## 2 RELATED WORK

*(a) Data Center Electricity Cost, Renewable Power, and Peak Demand:* There have been many recent efforts proposing methods to minimize electricity costs across geo-distributed data centers, with the fundamental decisions of the optimization problem relying on a TOU electricity pricing model [15], [16]. Electricity costs are often much higher during peak hours of the day (typically 8 A.M. to 5 P.M.). The electricity cost models, sometimes in combination with a model that considers renewable energy, motivates the use of optimization techniques to minimize energy cost or, if provided a revenue model, to maximize total profit [17], [18]. A few other papers consider peak demand pricing models often in conjunction with energy storage devices/batteries to optimize electricity costs [19], [20].

*(b) Data Center Network Management:* Many recent papers focus on intelligent workload scheduling among cloud centers to optimize the quality of service (QoS) while increasing cloud operating profits [21], [22]. As data migration among cloud data centers increases network costs, some efforts try to address this issue by proposing techniques that intelligently allocate workloads to reduce overall network costs [23], [24]. In [25] and [26], the authors propose a multi-objective framework that simultaneously optimizes the resource wastage and migration costs while meeting QoS requirements.

*(c) Game Theory for Data Center Resource Management:* A few efforts consider game-theoretic approaches for cost optimization in geo-distributed cloud data centers. In [27], the authors propose a technique to allocate computing resources according to the service subscriber's requirements by using non-cooperative game theory. The authors in [28], [29] propose a bandwidth resource management technique for geo-distributed cloud data centers. In [30], the authors propose an algorithm that tries to reduce data migration costs and maximize the profit of all cloud service providers involved in a federation. In [31], the authors use a cooperative game theory to model the cooperative electricity procurement process of data centers as a cooperative game and show the cost-saving benefits of aggregation. In [32], an approach is proposed to address the tradeoff between QoS and energy consumption in cloud data centers. The approach uses a game-theoretic formulation that allows for appropriate loss of QoS while maximizing the cloud service provider's profits. In [33], the authors consider a cloud data center and smart grid utility demand-response program. The work proposes a cooperative game theory-based technique to migrate workloads between data centers and better utilize the benefits provided by demand response schemes over multiple data center locations.

In [34] and [35], the authors propose a non-cooperative game-theoretic workload management technique to minimize data center energy costs, while considering TOU electricity pricing information.

Similar to [34] and [35], our work uses information about TOU electricity pricing and develops a non-cooperative game-theoretic workload management technique to minimize the cloud operating cost. Our framework also uses information about renewable power and net metering policies at various data center locations like [34]. However, both [34] and [35] do not consider detailed models for data center compute and cooling power, co-location interference, and peak demand pricing distribution. Moreover, these efforts do not consider heterogeneity in workload (applications/task types) executing in the data center) and heterogeneity within the data center (types of servers, server rack arrangements, etc.). Our previous work [36] considers many of the above-mentioned aspects. We extend [36] by developing a new inter-data center network cost model, a data center queueing delay model, and considering these factors in our resource management problem. Moreover, we propose a new game theory-based workload management framework that takes a holistic approach to the cloud operating cost minimization problem, which is shown to be more effective than the multiple resource management strategies presented in [36] (Section 7 presents a detailed analysis to quantify the improvement).

## 3 System Model

### 3.1 Overview

Our framework comprises a geo-distributed-level <u>C</u>loud <u>W</u>orkload <u>M</u>anager (CWM) that distributes incoming workload requests to geographically distributed data centers. Each data center has its own local <u>D</u>ata center <u>W</u>orkload <u>M</u>anager (DWM) that takes the workload assigned to it by the CWM and maps requests to compute nodes within the data center. We first describe the system model at the geo-distributed level and then provide further details into the models of components at the data center level. We provide a list of abbreviations and notations in the appendix.

### 3.2 Geo-Distributed Level Model

We consider a rate-based workload management scheme, where the workload arrival rate can be predicted over a decision interval called an epoch [13], [14]. In our work, an epoch length $T^e$ is one hour, and a 24-epoch period represents a full day. Within the short duration of an epoch, the workload arrival rates can be reasonably approximated as constant, e.g., the Argonne National Lab Intrepid log shows mostly constant arrival rates over larger intervals of time [37].

Let $D$ be a set of $|D|$ data centers and let $d$ represent an individual data center. We assume that a cloud infrastructure (Fig. 1) is composed of $|D|$ data centers, that is, $d = 1, 2, ..., |D|$ and $d \in D$. Let $I$ be a set of $|I|$ task types and $i$ represents an individual task type. We consider $|I|$ task (i.e., workload) types, that is, $i = 1, 2, ..., |I|$ and $i \in I$. A task type $i \in I$ is characterized by its arrival rate and its execution rate, i.e., reciprocal of the estimated time required to complete a task of task type $i$ on each of the heterogeneous compute nodes, in each performance state (P-state). We assume that the beginning of each epoch $\tau$ represents a steady-state scheduling problem where the CWM splits the global <u>a</u>rrival <u>r</u>ate $GAR_i(\tau)$ for each task type $i$ into the local data center level <u>a</u>rrival <u>r</u>ate $AR_{i,d}(\tau)$ and assigns it to each data center $d \in D$. That is,

$$\sum_{d=1}^{|D|} AR_{i,d}(\tau) = GAR_i(\tau), \quad \forall i \in I. \tag{1}$$

The CWM performs this assignment such that the total cloud operating (energy and network) cost across all data centers is minimized, with the constraint that the execution rates of all task types exceed their arrival rates at each data center, i.e., all tasks complete without being dropped or unexecuted. At the start of each epoch, the CWM calculates (explained in Section 3.3.2) a co-location aware data center maximum <u>ex</u>ecution <u>r</u>ate $ER_{i,d}$ for each task type $i$ at each data center $d$. Later, it splits global arrival rate $GAR_i(\tau)$ for each task type $i$ into local data center level arrival rates $AR_{i,d}(\tau)$ such that the data center maximum execution rate $ER_{i,d}(\tau)$ exceeds the corresponding arrival rate $AR_{i,d}(\tau)$, thus ensuring the workload is completed. That is,

$$ER_{i,d}(\tau) > AR_{i,d}(\tau), \quad \forall i \in I, \forall d \in D. \tag{2}$$

Note that $ER_{i,d}(\tau)$ must be strictly greater than $AR_{i,d}(\tau)$ to minimize the expected average queueing delay for task type $i$ at data center $d$, as discussed later in Section 3.3.8.

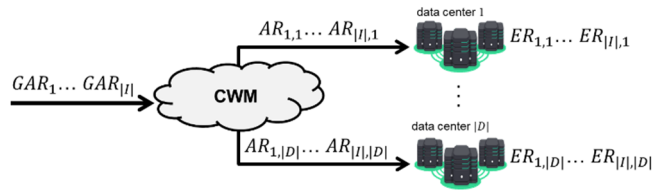

Fig. 1. Cloud workload manager (CWM) performing geo-distributed level task (workload) assignment to data centers

### 3.3 Data Center Level Model

#### 3.3.1 Organization of Each Data Center

Each data center $d$ houses $NN_d$ number of <u>n</u>odes and a cooling system comprised of <u>c</u>omputer <u>r</u>oom <u>a</u>ir <u>c</u>onditioning (CRAC) units. Let $NCR_d$ be the <u>n</u>umber of CRAC units. Heterogeneity exists across compute nodes, where nodes vary in their execution speeds, power consumption characteristics, and the number of cores. The <u>n</u>umber of <u>c</u>ores in <u>n</u>ode $n$ is $NCN_n$, and $NT_k$ is the <u>n</u>ode <u>t</u>ype to which core $k$ belongs.

#### 3.3.2 Co-Location Aware Execution Rates

Tasks competing for shared memory in multicore processors can cause severe performance degradation, especially when competing tasks are memory intensive. The memory-intensity of a task refers to the ratio of last-level cache misses to the total number of instructions executed. We use a linear regression model that combines a set of disparate features based on the current tasks assigned to a multicore processor to predict the execution time of a target task $i$ on core $k$ in the presence of performance degradation due to interference from task co-location [38].

We classify the task types into memory-intensity classes on each of the node types, and calculate the coefficients for each memory-intensity class using the linear regression model to determine a co-located execution rate for task type $i$ on core $k$, $CER_{i,k}^{core}(\tau)$ [38]. When considering co-location at a data center $d$, the co-location aware data center execution rate for task type $i$ is given by:

$$ER_{i,d}(\tau) = \sum_{n=1}^{NN_d} \sum_{k=1}^{NCN_n} CER_{i,k}^{core}(\tau). \qquad (3)$$

The linear regression model was trained using execution time data that was collected by executing benchmarks from the BigDataBench 5.0 [39] benchmark suite on a set of server-class multicore processors that define the nodes used in our study (see Table 1 in Section 6.2 for node type details). This model for execution time prediction under co-location interference is derived from real workloads and real server machines and has a mean prediction error of approximately 7% [38].

### 3.3.3 Data Center Power Model

We use the detailed data center power model from our prior work [36]. Let $PCR_{d,c}(\tau)$ be the power consumed by a CRAC unit $c$ in data center $d$. $PN_n(\tau)$ is the power consumption of node $n$. Each data center is partially powered by either solar power, wind power, or some combination of both. The total renewable power, $PR_d(\tau)$, available at data center $d$ is the sum of the wind and solar power available at that time. $PR_d(\tau)$ can be zero if no renewable power is available. Let $Eff_d$ be an approximation of the power overhead coefficient in data center $d$ due to the inefficiencies of power supply units. $Eff_d$ is always greater than or equal to 1. The net total power consumed throughout data center $d$, $PD_d(\tau)$, is calculated as:

$$PD_d(\tau) = \left( \sum_{c=1}^{NCR_d} PCR_{d,c}(\tau) + \sum_{n=1}^{NN_d} PN_n(\tau) \right) \cdot Eff_d - PR_d(\tau). \quad (4)$$

For epoch $\tau$, $PD_d(\tau)$ can be negative if the renewable power available at the data center, $PR_d(\tau)$, is greater than the cooling and computing power.

### 3.3.4 Net Metering Model

Net metering allows data center operators to sell back the excess renewable power generated on-site to the utility company. When the excess power is added into the grid, utility companies pay a fraction of the retail price. The value of this factor varies by data center location. This fraction is called the net metering factor, $\alpha_d$.

### 3.3.5 Peak Demand Model

Most utility providers charge a flat-rate (peak demand) fee based on the highest (peak) power consumed at any instant during a given billing period, e.g., month. The peak demand price per $kW$ at data center $d$ is denoted as $P_d^{price}$. We define $Pc_d^{peak}(\tau)$ as the highest grid power consumed since the beginning of the current month, including the current epoch $\tau$. We define $Pp_d^{peak}(\tau)$ as the highest grid power consumption since the beginning of the current month until the start of the current epoch $\tau$. The peak power cost increase at data center $d$, $\Delta_d^{peak}(\tau)$, is then defined as:

$$\Delta_d^{peak}(\tau) = P_d^{price} \cdot \left( Pc_d^{peak}(\tau) - Pp_d^{peak}(\tau) \right) \qquad (5)$$
if $Pc_d^{peak}(\tau) \geq Pp_d^{peak}(\tau)$, else it is equal to 0.

The peak power cost increase $\Delta_d^{peak}(\tau)$ is calculated in each epoch $\tau$ and summed over all epochs in a billing period to calculate total peak power cost.

### 3.3.6 Network Cost Model

Two principal components of the network cost are the network price per data traffic unit ($/GB), $N^{price}$, and the amount of data volume (GB) for the number of tasks migrated (outward). We assume that $N^{price}$ is same across all data centers. Let $S_i$ be the size (in GB) of a task type $i$. The network cost at data center $d$, $NC_d(\tau)$, is calculated as

$$NC_d(\tau) = \sum_{i=1}^{|I|} \sum_{n=1}^{NN_d} (N^{price} \cdot S_i). \qquad (6)$$

### 3.3.7 Data Center Cost Model

The electricity price per $kWh$ at data center $d$ is defined as $E_d^{price}(\tau)$. Data center operators can use net metering if the net total power consumed throughout the data center, $PD_d(\tau)$, is negative. For such conditions, the total data center cost $DC_d(\tau)$ for data center $d$ can be defined as

$$DC_d(\tau) = E_d^{price}(\tau) \cdot \alpha_d \cdot PD_d(\tau) + \Delta_d^{peak}(\tau) + NC_d(\tau) \qquad (7)$$

where $\alpha_d = 1$ if $PD_d(\tau)$ is positive and $0 \leq \alpha_d \leq 1$ otherwise. The first term in (7) represents the TOU electricity cost, the second term represents the peak demand cost, and the third represents the network cost.

### 3.3.8 Data Center Queueing Delay Model

To reduce operating costs, cloud service providers tend to allocate more workload to the more energy-efficient data centers, often running them at very high capacity. In some cases, this can overwhelm the data center servers and network equipment causing a queueing delay [40]. This manifests as delay experienced by users of the cloud service. Because of queueing delay, data centers fail to service the incoming workload in a timely manner, which causes revenue loss. There is evidence that lost revenue has a linear relationship to the data center queueing delay for sites such as Google, Bing, and Shopzilla [41]. These assumptions are satisfied by most standard queueing methods, e.g., the 95th percentile of delay under the M/M/1 queue and the average delay under the M/GI/1 processor sharing (PS) queue [35] and [40]. Let $d_{i,d}^{queue}(\tau)$ be the expected average queueing delay for the task type $i$ at data center $d$, is calculated as:

$$d_{i,d}^{queue}(\tau) = \frac{1}{ER_{i,d}(\tau) - AR_{i,d}(\tau)}. \qquad (8)$$

## 4 PROBLEM FORMULATION

We consider a scenario with multiple data centers maintained by a cloud service provider. The data centers can share the workload coming in from various locations. The system is assumed to be under-subscribed in the sense that the system is expected to have enough computation resources to complete the workload without requiring that any tasks be dropped or terminated before completion. The tasks originate off-site from the data centers, and we do not consider the transmission time and cost from a task origin to a data center. If the CWM migrates a task from one location to another location, there is a data transfer cost (as discussed in Section 3.3.6) associated with it. The objective of a CWM is to allocate the workload across geo-distributed data centers to minimize the monetary cloud operating (energy and network) cost of the system (the sum of (7) across all data centers) while ensuring that the workload is completed according to the constraint defined by (2).

The complexity of the CWM workload allocation problem makes it NP-hard [42], and therefore we propose a game theory-based workload management heuristic. Here, we model the cloud workload distribution problem as a non-cooperative game. In a non-cooperative game, there could be a finite (or infinite) number of players who aim to maximize/minimize their objective independently but ultimately reach an equilibrium. For a finite number of players, this equilibrium is called the Nash equilibrium [43].

# 5 NASH EQUILIBRIUM-BASED HEURISTIC

## 5.1 Overview

Our proposed Nash equilibrium-based intelligent load distribution (NILD) heuristic is a non-cooperative game-theoretic load balancing approach. The following subsections describe the components of the heuristic in more detail.

## 5.2 Objective Function

NILD jointly optimizes the estimated data center (energy and network) cost and the estimated delay cost, as discussed in the following subsections.

### 5.2.1 Estimated Data Center Cost

Let $J_d$ be a set of node types in a data center $d$ and $j$ represents an individual node type. In a data center $d$, there are a total of $NN_{d,j}$ nodes of node type $j$. Let $P_j^D$ be the average peak dynamic power for node type $j$. It is calculated by averaging (over all task types) the peak power for each task type $i$ executing on node type $j$. Let $PD_d^{max}$ be the maximum power dissipation possible at data center $d$, is calculated as:

$$PD_d^{max} = \left( NCR_d \cdot PCR_{d,c}^{max} + \sum_{j \in J_d} NN_{d,j} \cdot P_j^D \right) \cdot Eff_d. \quad (9)$$

Recall that $PR_d(\tau)$ is the total renewable power available at data center $d$. Let $PD_{i,d}^E(AR, \tau)$ be the estimated power dissipation possible for each task type $i$ at data center $d$, is calculated as:

$$PD_{i,d}^E(AR, \tau) = \left( \frac{PD_d^{max} \cdot AR_{i,d}(\tau)}{ER_{i,d}(\tau)} - PR_d(\tau) \right). \quad (10)$$

Let $NC_{i,d}^{max}$ be the maximum network cost possible for each task type $i$ at data center $d$, is calculated as:

$$NC_{i,d}^{max} = N^{price} \cdot NN_d \cdot S_i. \quad (11)$$

Let $NC_{i,d}^E(AR, \tau)$ be the estimated network cost possible for each task type $i$ at data center $d$, is calculated as:

$$NC_{i,d}^E(AR, \tau) = \frac{NC_{i,d}^{max} \cdot AR_{i,d}(\tau)}{ER_{i,d}(\tau)}. \quad (12)$$

Observe that, at data center $d$ with zero $PR_d(\tau)$ for each task type $i$, $PD_{i,d}^E(AR, \tau)$ in (10) and $NC_{i,d}^E(AR, \tau)$ in (12) will increase to $PD^{max}$ and $NC_{i,d}^{max}$, respectively as the data center arrival rates $AR_{i,d}(\tau)$ approaches to its maximum execution rate $ER_{i,d}(\tau)$. Thus, we can say that both $PD_{i,d}^E(AR, \tau)$ and $NC_{i,d}^E(AR, \tau)$ are functions of $AR$ and $\tau$.

Then the estimated data center cost $DC_{i,d}^E(AR, \tau)$ incurred by the task type $i$ with a data center maximum execution rate $ER_{i,d}(\tau)$ at data center $d$, can be calculated by modifying (7) as:

$$DC_{i,d}^E(AR, \tau) = E_d^{price}(\tau) \cdot \alpha_d \cdot PD_{i,d}^E(AR, \tau) + \Delta_d^{peak}(\tau) + NC_{i,d}^E(AR, \tau) \quad (13)$$

where $\alpha_d = 1$ if $PD_{i,d}^E(AR, \tau)$ is positive and $0 \leq \alpha_d \leq 1$ otherwise.

### 5.2.2 Estimated Delay Cost

The estimated delay cost is associated with the data center queueing delay (8) and captures the lost revenue incurred from the delay experienced by the requests. For this loss, we use the model from [35]. NILD considers the queueing delay within the data center. Let $\beta$ be a constant known as a delay cost factor. As per (8), the estimated delay cost $DelC_{i,d}^E(AR, \tau)$ for the task type $i$ at the data center $d$, can be calculated as

$$DelC_{i,d}^E(AR, \tau) = \beta \cdot AR_{i,d}(\tau) \cdot d_{i,d}^{queue}(\tau). \quad (14)$$

Substituting the value of $d_{i,d}^{queue}(\tau)$ from (8), we get:

$$DelC_{i,d}^E(AR, \tau) = \frac{\beta \cdot AR_{i,d}(\tau)}{ER_{i,d}(\tau) - AR_{i,d}(\tau)}. \quad (15)$$

Observe that $DelC_{i,d}^E(AR, \tau)$ will increase (to infinity) as the difference $(ER_{i,d}(\tau) - AR_{i,d}(\tau))$ decreases to zero. This shows that the data centers running at very high capacity result in a very high estimated delay cost. Because of this, NILD avoids assigning the highest possible arrival rates to data centers. Therefore, with a constant $ER_{i,d}(\tau)$ and $\beta$, the estimated delay cost becomes a function of $AR$ and $\tau$.

### 5.2.3 Estimated Overall Cost Incurred

The goal of NILD is to minimize the overall cost incurred, which has two components: estimated data center cost, defined in (13), and the estimated delay cost, defined in (15). The estimated overall cost $OC_i^E(AR, \tau)$ incurred for the task type $i$, can be calculated as:

$$OC_i^E(AR, \tau) = \sum_{d=1}^{|I|} \left( DC_{i,d}^E(AR, \tau) + DelC_{i,d}^E(AR, \tau) \right). \quad (16)$$

## 5.3 Load Distribution as a Non-cooperative Game

In our workload distribution problem, the objective of NILD is to split the global arrival rate $GAR_i(\tau)$ for each task type $i$ into local data center level arrival rates $AR_{i,d}(\tau)$ and assign it to each data center $d \in D$. We model this problem as a non-cooperative game that is played among a set of players. Each player has a set of strategies and the estimated overall cost that results from using each strategy. In this game, each task type $i$ is a player. NILD uses Algorithm 1, as discussed later in Section 5.5, to find the strategy $AR_i$ of each player $i$. Then NILD uses Algorithm 2, as discussed later in Section 5.6, to find a feasible cloud workload distribution strategy $AR = \{AR_1, AR_2, \ldots, AR_{|I|}\}$ such that $OC_i^E(AR, \tau)$ is minimized. The components of our non-cooperative game are:

- **Players:** a finite set of players, where each task type $i$ is a player; $i = 1, 2, \ldots, |I|$ (as per Section 3.2)
- **Strategy sets:** load distribution strategy of a player $AR_i = \{AR_{i,1}, AR_{i,2}, \ldots, AR_{i,|D|}\}$
- **Cost:** each player wants to minimize the overall cost $OC_i^E(AR, \tau)$ associated with its strategy.

### 5.4 Nash Equilibrium

To obtain the workload distribution strategy for the cloud data centers, the game described in Section 5.3 can be solved using a Nash equilibrium, which is the most widely used solution for such games. The Nash equilibrium of our non-cooperative game is a load distribution strategy of the entire cloud $AR = \{AR_1, AR_2, \ldots, AR_{|I|}\}$ such that for each player $i$:

$$AR_i \in \underset{AR_i}{arg\,min}\, OC_i^E(AR, \tau). \qquad (17)$$

If no player can further minimize the overall cost incurred by changing its current strategy to another one, the strategy $AR$ is a Nash equilibrium. In this equilibrium, a player $i$ cannot decrease the overall cost incurred by choosing a different workload distribution strategy $AR_i$ given the other players' workload distribution strategies.

### 5.5 Best Reply Strategy

In the Nash equilibrium, the strategy profile $AR$ is such that each player's workload distribution strategy is the best reply given to the other players' strategies. The best reply for a player provides a minimum overall cost for that player's workload allocation strategy given the other players' strategies. Therefore, by finding the best reply strategy $AR_i$ for each player $i$ we determine the load distribution strategy of the entire cloud $AR = \{AR_1, AR_2, \ldots, AR_{|I|}\}$. The NILD assigns arrival rates to a data center incrementally. After each iteration, the available capacity (maximum execution rate) of a data center decreases. This leftover capacity (execution rate) is known as the available execution rate. Let $ER_{i,d}^{av}$ be the available execution rate for a task type player $i$ at data center $d$, is calculated as

$$ER_{i,d}^{av}(\tau) = ER_{i,d}(\tau) - AR_{i,d}(\tau). \qquad (18)$$

The problem of finding the best reply strategy depends on finding the optimal workload distribution for a system with each player, i.e., task type $i$ ($\forall i \in I$), and $|D|$ distributed data centers with available execution rates $ER_{i,d}^{av}$ ($\forall d \in D$). We can represent the above problem in the following optimization equation to find the best reply:

$$\underset{AR_i}{min}\, OC_i^E(AR, \tau) \qquad (19)$$

subject to the constraints defined by (1) and (2).

The decision variable involved in this optimization is $AR_i = \{AR_{i,1}, AR_{i,2}, \ldots, AR_{i,|D|}\}$, because we assume that, at equilibrium, the strategies of other players are constants. As our optimization framework has the objective of minimizing the estimated operating costs, we try to assign larger workloads to less costly data centers. We determine whether a data center is expensive by calculating the value for a metric called capacity factor [35]. A lower capacity factor represents a less expensive data center, and vice versa. The capacity factor is nothing but the estimated data center maximum cost that is independent of the arrival rate. Let $CF_{i,d}^E(\tau)$ be the estimated capacity factor of a data center $d$ for task type $i$. Modifying (13) we get:

$$CF_{i,d}^E(\tau) = E_d^{price}(\tau) \cdot \alpha_d \cdot \left( \frac{PD_d^{max}}{ER_{i,d}^{av}(\tau)} - PR_d(\tau) \right)$$
$$+ \Delta_d^{peak}(\tau) + \left( \frac{NC_{i,d}^{max}}{ER_{i,d}^{av}(\tau)} \right) \qquad (20)$$

where $\alpha_d = 1$ if $PD_d^{max}$ is positive and $0 \leq \alpha_d \leq 1$ otherwise.

First sort the data centers based on their estimated capacity factors $CF_{i,d}^E(\tau)$. As per [35], let $q$ be the smallest integer (number of data centers) that satisfies

$$\sum_{d=1}^{q} \sqrt{\frac{\beta \cdot ER_{i,d}^{av}(\tau)}{\lambda(\tau) - CF_{i,d}^E(\tau)}} < \sum_{d=1}^{q} ER_{i,d}^{av}(\tau) - GAR_i(\tau). \qquad (21)$$

where $\lambda(\tau)$ is a Lagrangian multiplier [35], whose value is given by modifying (20) as:

$$\lambda(\tau) = E_q^{price}(\tau) \cdot \alpha_q \cdot \left( \frac{PD_q^{max}}{ER_{i,q}^{av}(\tau)} - PR_q(\tau) \right)$$
$$+ \Delta_d^{peak}(\tau) + \left( \frac{NC_{i,q}^{max}}{ER_{i,q}^{av}(\tau)} \right) + \frac{\beta}{ER_{i,q}^{av}(\tau)} \qquad (22)$$

where $\alpha_d = 1$ if $PD_d(\tau)$ is positive and $0 \leq \alpha_d \leq 1$ otherwise.

To further simplify (21), let $\gamma$ be the right side and $\delta$ be the left side of (21), their values are given by:

$$\gamma = \sum_{d=1}^{q} ER_{i,d}^{av}(\tau) - GAR_i(\tau). \qquad (23)$$

$$\delta = \sum_{d=1}^{q} \sqrt{\frac{\beta \cdot ER_{i,d}^{av}(\tau)}{\lambda(\tau) - CF_{i,d}^E(\tau)}}. \qquad (24)$$

Therefore, (21) can be rewritten as:

$$\delta < \gamma \qquad (25)$$

As per [35], the arrival rate $AR_{i,d}(\tau)$ for a player $i$ at data center $d$ can be calculated as:

$$AR_{i,d}(\tau) = ER_{i,d}^{av}(\tau) - \frac{\gamma}{\delta} \cdot \sqrt{\frac{\beta \cdot ER_{i,d}^{av}(\tau)}{\lambda(\tau) - CF_{i,d}^E(\tau)}} \qquad (26)$$

if $1 \leq d < q$ else it is equal to 0.

The constraint (25) ensures that the workload is distributed starting from the least expensive data center first to the expensive one at the end. For the data center with the lowest $CF_{i,d}^E$, the second term in (26) will be the lowest, assigning larger $AR_{i,d}$ to inexpensive data centers, and vice versa. Equation (26) is used to find the best reply strategy $AR_i = \{AR_{i,1}, AR_{i,2}, \dots, AR_{i,|D|}\}$.

Based on (25) and (26), the following *Best-Reply* algorithm is formulated to find the optimal load distribution strategy, i.e., the best reply of player $i$.

---

**Algorithm 1.** Pseudo-code for *Best-Reply* strategy

---

**inputs**: global arrival rate: $GAR_i(\tau)$
available execution rates: $ER_{i,1}^{av}(\tau), \dots, ER_{i,|D|}^{av}(\tau)$
capacity factors: $CF_{i,1}^E(\tau), \dots, CF_{i,|D|}^E(\tau)$
**output**: load distribution strategy for a player :
    $AR_i = \{AR_{i,1}, \dots, AR_{i,|D|}\}$
1.  sort data centers in ascending order of capacity factors
    i.e., $CF_{i,1}^E(\tau) \leq \dots \leq CF_{i,d}^E(\tau) \leq \dots \leq CF_{i,|D|}^E(\tau)$
2.  $q = |D|$
3.  initialize $\gamma$ and $\delta$ using (23) and (24), respectively
4.  **while** $\gamma > \delta$
5.      $\gamma = \gamma - ER_{i,d}^{av}(\tau)$
6.      $AR_{i,q} = 0$
7.      $q = q - 1$
8.      update $\delta$ using (24)
9.  **for** each data center in $\{1,2,\dots,q\}$
10.     calculate $AR_{i,d}(\tau)$ using (26)

---

The *Best-Reply* algorithm takes global arrival rate, available execution rates, and capacity factors of player $i$ as inputs. As the output, it determines the best reply strategy $AR_i = \{AR_{i,1}, AR_{i,2}, \dots, AR_{i,|D|}\}$. First, the data centers are sorted based on their capacity factors (step 1). The heuristic aims to assign larger workloads to less expensive data centers. It initializes the values of $q$, $\gamma$, and $\delta$ (steps 2 and 3). We use the while loop (steps 4-8) to find the smallest index data center that satisfies (25). In the loop, the algorithm does not assign any workload to the last (most expensive) data center i.e., $AR_{i,q} = 0$, if the $(q-1)^{th}$ data center's available execution rate is capable of satisfying (25). This while loop updates the values of $q$, $\gamma$, and $\delta$ and continues until the condition (25) or $\gamma > \delta$ is met. Finally, the algorithm determines $AR_{i,d}(\tau) \in AR_i$ for each data center using (26).

### 5.6 NILD Heuristic

We designed the final NILD algorithm to compute the Nash equilibrium of the non-cooperative game. It uses the *Best-Reply* algorithm explained in the previous section. The CWM uses Algorithm 2 shown below to assign (i.e., split) the global workload arrival rates to data center level arrival rates for all task types. Each player $i$ computes its optimal *Best-Reply* strategy (steps 3-8), i.e., it determines $AR_{i,d}(\tau)$ for each data center $i$. The heuristic then calculates the *norm*, by summing the absolute value of the difference among its operating costs across all players (task types) from the current $l^{th}$ iteration and the previous $(l-1)^{th}$ iteration. This continues until the algorithm comes to an equilibrium i.e., the difference in the total operating cost across all the players in successive iterations (i.e., *norm*) is less than the pre-defined stopping criterion (tolerance $\varepsilon$). The value of $\varepsilon$ is determined empirically through simulation studies to provide a value that gives the system the best possible performance. Here, each player determines its *Best-Reply* strategy using the current load distribution strategies of other players and updates its strategy.

---

**Algorithm 2.** Pseudo-code for NILD heuristics

---

**inputs**: global arrival rates:
    $GAR_1(\tau), \dots, GAR_i(\tau), \dots, GAR_{|I|}(\tau)$
data center maximum execution rates:
    $ER_{1,1}(\tau), \dots, ER_{i,d}(\tau), \dots, ER_{|I|,|D|}(\tau)$
**output**: cloud load distribution strategy:
    $AR = \{AR_1, \dots, AR_i, \dots, AR_{|I|}\}$
1.  **initialize:**
    iteration number: $l = 0$
    load distribution strategy: $AR_i = 0$
    estimated overall cost incurred: $OC_i^E(AR, \tau) = 0$
    tolerance: $\varepsilon = 10^{-3}$
    $flag = continue$
2.  **while** $flag = continue$
3.    **for** each $i$ in $\{1,2,\dots,|I|\}$
4.      **for** each $d$ in $\{1,2,\dots,|D|\}$
5.        calculate $ER_{i,d}^{av}(\tau)$ using (18)
6.        calculate $CF_{i,d}^E(\tau)$ using (20)
7.      find optimal $AR_i$ using *Best-Reply* algorithm
8.      calculate $OC_i^E(AR, \tau)$ using (16)
9.    $norm = \sum_{i=1}^{|I|} |OC_i^{E(l-1)}(AR, \tau) - OC_i^{E(l)}(AR, \tau)|$
10.  **if** $norm < \varepsilon$
11.    $flag = stop$

---

## 6 SIMULATION ENVIRONMENT
### 6.1 Comparison Heuristics

We compare our proposed NILD heuristic with (a) co-location aware force-directed load distribution (FDLD) [36], (b) genetic algorithm load distribution (GALD) [36], and (c) Nash equilibrium based simple load distribution (NSLD) [35]. FDLD [36] is a variation of force-directed scheduling [44], frequently used for optimizing semiconductor logic synthesis. FDLD is an iterative heuristic that selectively performs operations to minimize system forces until all constraints are met. The Genitor style [45] GALD has two parts: a genetic algorithm-based CWM and a local data center level greedy heuristic that is used to calculate the fitness value of the genetic algorithm. The local greedy heuristic has information about task-node power dynamic voltage and frequency scaling (DVFS) models [36]. [35] also proposes a game-theoretic formulation for the load distribution problem. However, it addresses a much simpler problem, with homogeneous workloads and datacenters, and a simplistic data center cost model. We adapt the approach proposed in [35] to our heterogeneous environment we have outlined in Section 3, and create the NSLD. The simpler models used NSLD affect the data center and delay costs. They also affect the data center capacity factor and arrival rate calculation, which changes the *Best-Reply* strategy of a player (task type). In contrast, our proposed NILD framework integrates detailed models for data center compute and cooling power, co-location interference, net metering, peak demand, and network pricing distribution. Moreover, we also consider heterogeneity in workload (appli-cations/task types executing in the data center) and heterogeneity within the data center (types of servers, server rack arrangements, etc.), with more detailed models to calculate the latency and power overhead for computation and communication.

## 6.2 Experimental Setup

Experiments were conducted for three geo-distributed data center configurations containing four, eight, and sixteen data centers. Locations of the data centers in the three configurations were selected from major cities around the continental United States to provide a variety of wind and solar conditions among sites and at various times of the day (Fig. 2). The sites of each configuration were selected so that each configuration would have an even east coast to west coast distribution to better exploit TOU pricing, peak demand pricing, net metering, and renewable power. Each data center comprises 4,320 nodes arranged in four aisles, and is heterogeneous within itself, having nodes from either two or three of the node types given in Table 1, with most locations having three node types and per-node core counts that range from 4 to 12 cores.

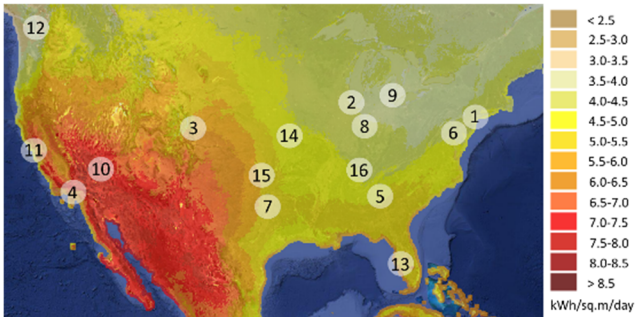

Fig. 2. Location of simulated data centers overlaid on solar irradiance intensity map (average annual direct normal irradiance [46])

TABLE 1: Node Processor Types Used in Experiments

| Intel processor | # cores | L3 cache | frequency range |
|---|---|---|---|
| Xeon E3-1225v3 | 4 | 8MB | 0.8 - 3.20 GHz |
| Xeon E5649 | 6 | 12MB | 1.60 - 2.53 GHz |
| Xeon E5-2697v2 | 12 | 30MB | 1.20 - 2.70 GHz |

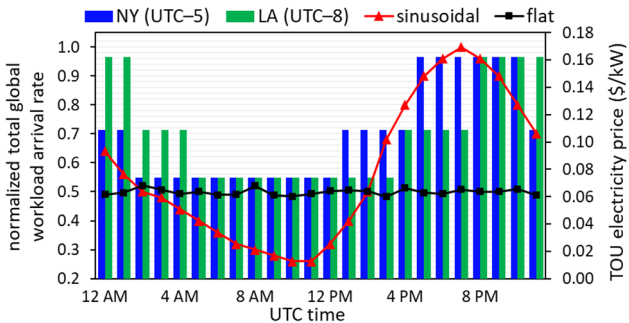

Fig. 3. Baseline task arrival rate and TOU prices at two locations (New York, Los Angeles) over 24 hours

The delay cost factor ($\beta$) used during NILD allocations was determined empirically (see Section 7.4.2) and set to 0.1. The time of each epoch $\tau$ was set to be one hour. The network prices used in the experiments were taken from Amazon Web Services [47]. The TOU electricity prices and the peak demand prices were taken directly from the utility. We assume that each data center has peak renewable power generating capacity equivalent, or slightly more at some locations, to its maximum power consumption. Solar and wind data were obtained from the National Solar Radiation Database [46]. At most locations, the net metering factor $\alpha_d$ is 1; in rare cases, it is less than 1; and in some cases, it is 0, i.e., net metering is not available at that location [12], [48].

Organizations are heavily using platform as a service (PaaS) offerings from cloud providers. These services mainly involve data warehouse, data analytics, and machine learning/artificial intelligence workloads [49], [50]. Due to the popularity of such workloads among cloud providers, we use data intensive (e.g., offline analytics and artificial intelligence) workloads from the BigDataBench 5.0 [39] benchmark suite. Table 2 summarizes the workloads from this suite that we consider in our work. Task execution times and co-located performance data for tasks of the different memory intensity classes were obtained from running the benchmark applications on the nodes listed in Table 1 [38]. Fig. 3 depicts a synthetic sinusoidal task arrival rate pattern and a flat task arrival rate pattern with line plots. The left y-axis is normalized to the highest total global workload arrival rate. A sinusoidal task arrival rate pattern exists in environments where the workload traffic depends on consumer interaction and follows their demand during the day, e.g., Netflix [51], Facebook [52]. However, for the environments where continuous computation is needed and the workload pattern is non-user/consumer interaction specific, the task arrival rate pattern is usually flat (nearly constant). Examples of such environments exist in military computing installations (Department of Defense) and government research labs (National Center for Atmospheric Research). For reference, Fig. 3 also depicts TOU prices for New York (east coast) and Los Angeles (west coast) data center locations with bar plots.

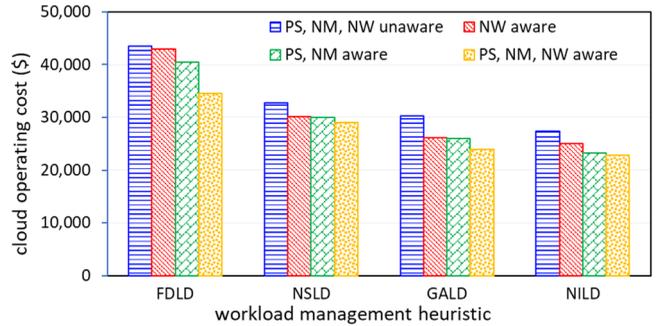

Fig. 4. Cloud operating costs for each heuristic over a day, for a configuration with four data centers

TABLE 2: Task Types Used in Experiments

| benchmark | workload type | dataset | size |
|---|---|---|---|
| LDA | offline analytics | Wikipedia entries | 233 MB |
| K-means | offline analytics | Facebook network | 650 MB |
| Naive Bayes | offline analytics | Amazon reviews | 500 MB |
| Image-to-Text | artificial intelligence | Microsoft COCO | 600 MB |
| Image-to-Image | artificial intelligence | cityscapes | 100 MB |

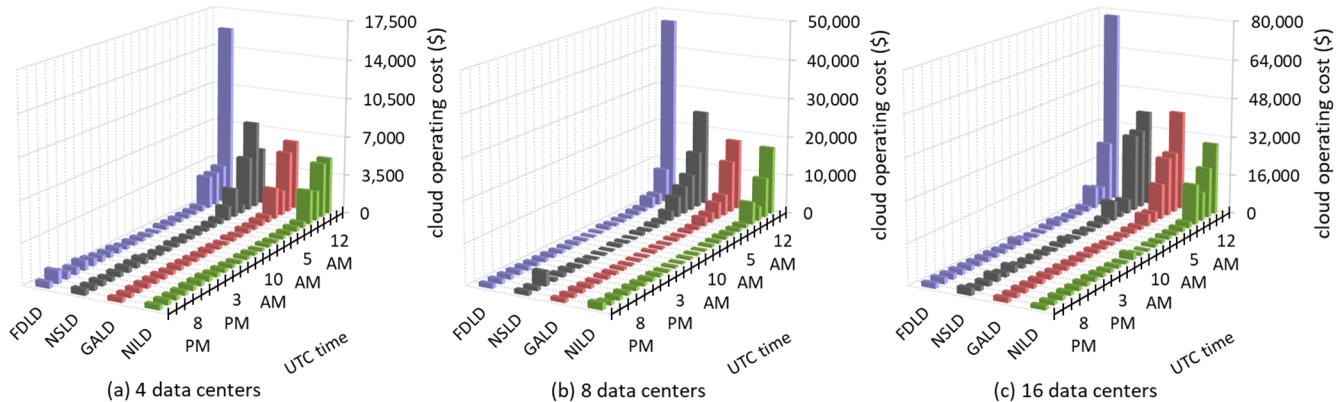

Fig. 5. Cloud operating costs for each heuristic over a day, for a configuration with (a) four, (b) eight, and (c) sixteen data centers

## 7 EXPERIMENTS

### 7.1 Cost Comparison of Heuristics

Our first set of experiments analyzes the total system energy cost for each heuristic described in Section 6.1. For each heuristic, we consider four variants: (a) peak shaving, net metering, and network (PS, NM, NW) unaware; (b) only network (NW) aware; (c) only peak shaving and net metering (PS, NM) aware; and (d) peak shaving, net metering, and network (PS, NM, NW) aware. Heuristic variants that are referred to as "peak shaving unaware" or "network unaware" do not include the peak demand pricing factor or the network cost, respectively, in their objective functions but consider them while calculating the total monthly operating cost at the end of the billing period. This experiment used a data center configuration with four locations running a sinusoidal workload pattern. The cloud operating costs were calculated over a duration of a day. The results are shown in Fig. 4.

For each heuristic; the 'PS, NM, NW aware' variant produced the best results. This validates our consideration of peak shaving, net metering, and network cost during the cloud workload distribution to more effectively minimize overall energy costs. It can also be observed that the FDLD heuristic performed the worst, severely over-provisioning nodes, and resulting in high operating costs. The GALD heuristic has information about task-node power (DVFS) models [36], allowing it to make better task placement decisions than FDLD. The NSLD [35] heuristic considers a simplistic view of data center compute and cooling power, while also ignoring co-location interference, net metering, and peak demand pricing distributions, and because it was designed for a different system model it is unable to perform as well as our NILD framework. The performance of the 'PS, NM aware' variant of NILD heuristic came close to its 'PS, NM, NW aware' variant. The 'PS, NM, NW aware' variant of NILD heuristic outperformed all other approaches. This heuristic minimizes the operating costs for all players/task types independently but ultimately reaches an equilibrium. Because of the non-cooperative nature of this method, each player/task type determines the lowest possible operating cost with its workload allocation strategy while considering the current load distribution strategies of other players.

### 7.2 Data Center Scalability Analysis
#### 7.2.1 Cost Reduction Comparison

In this experiment, we analyze heuristic performance for larger problem sizes. Simulations with the sinusoidal workload patterns were analyzed for eight and sixteen data center configurations in addition to the previously discussed four data center configuration. For each configuration, the percentage performance improvement of each heuristic over the 'PS, NM, NW unaware' FDLD variant is given in Table 3.

TABLE 3: Cloud Operating Cost Reduction Comparison

|  |  | heuristic | PS, NM, NW unaware | NW aware | PS, NM aware | |
|---|---|---|---|---|---|---|
| 4 data centers | FDLD | 0.0% | 1.3% | 6.9% | 20.5% | |
|  | NSLD | 24.7% | 30.8% | 31.1% | 33.1% | |
|  | GALD | 30.3% | 39.8% | 40.3% | 44.8% | |
|  | NILD | 37.0% | 42.5% | 46.5% | 47.5% | |
| 8 data centers | FDLD | 0.0% | 0.6% | 23.5% | 31.0% | |
|  | NSLD | 27.1% | 29.1% | 29.5% | 31.4% | |
|  | GALD | 33.7% | 32.7% | 48.6% | 49.2% | |
|  | NILD | 47.0% | 50.5% | 53.8% | 54.3% | |
| 16 data centers | FDLD | 0.0% | 0.4% | 22.6% | 34.7% | |
|  | NSLD | 32.6% | 34.5% | 34.7% | 36.8% | |
|  | GALD | 31.7% | 22.4% | 40.3% | 40.4% | |
|  | NILD | 46.9% | 50.6% | 50.5% | 54.2% | |

For GALD, going from 8 to 16 data centers, we notice that the energy cost reduction decreases with the increasing number of data centers. Here, as the number of data centers in the group grows larger, the problem size increases, and the number of GALD generations that can take place within the time limit (one hour by default) decreases, which decreases the performance of GALD. In contrast, FDLD, NSLD, and NILD reach the solution within minutes or seconds. These experiments confirm that the NILD heuristic consistently performs the best for all problem sizes. In case of the data center configurations containing sixteen data centers, the 'PS, NM, NW aware' NILD heuristic (our proposed work) achieved 41% and 23% better cost savings than the 'PS, NM aware' FDLD and GALD heuristics (from our previous work [36]), respectively. It also achieved 32% lower cost savings than the 'PS, NM, NW unaware' NSLD heuristic (from the previous work [35]).

### 7.2.2 Epoch based Analysis

For most of our experiments, we analyzed the total system cost for each heuristic over one day. Fig. 5 shows a detailed view of the cloud operating cost at one-hour intervals over a day for four, eight and sixteen data centers executing a sinusoidal workload. We consider the 'PS, NM, NW aware' variants of all workload management heuristics in this study. The operating cost for each heuristic is very high during the first epoch in the figure because the period for which the results are shown represents the first day of the month where the initial peak demand cost is added. This effect stays there for the first day and would not be present for other days of the month. After a few epochs, the performance of the GALD came close to the NILD but could not surpass its performance in general.

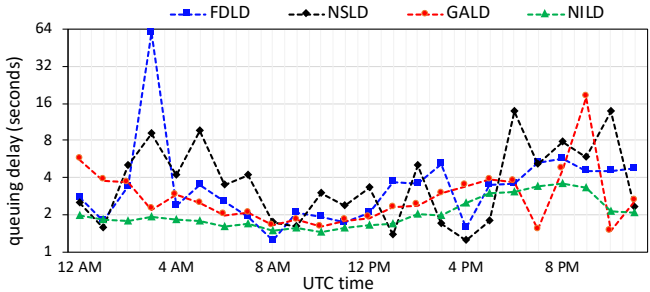

Fig. 6. Data center queueing delay comparison among heuristics over a day executing sinusoidal workload for eight data centers

We discussed the notion of the data center queueing delay in Section 3.3.8. It is calculated using (8). We also discussed how important it is for cloud service providers to minimize this delay. For this experiment, we considered the 'PS, NM, NW aware' variant of each heuristic for a group of eight data centers executing sinusoidal workload over a day. In Fig. 6, the colored data points represent the queueing delay (summed over 8 data centers) values for all heuristics for each epoch over a day. As per (8), the queueing delay increases as the denominator's value (the difference $ER_{i,d}(\tau) - AR_{i,d}(\tau)$') decreases. The increase in the queueing delay across heuristics indicates the times during which the data centers are running at very high or nearly maximum capacity ($ER_{i,d}(\tau)$) due to the resource allocation decisions ($AR_{i,d}(\tau)$) made by the heuristics. The results shown in Fig. 6 reveal that the FDLD, NSLD, and GALD heuristics fail to keep the delay small for certain epochs, whereas the NILD heuristic maintains a lower queueing delay compared to the other heuristics over the day (on average) across epochs.

TABLE 4: Heuristic Runtime Comparison

| # data centers | FDLD runtime | NSLD runtime | NILD runtime |
|---|---|---|---|
| 4 | ~1 min | ~1 seconds | ~2 seconds |
| 8 | ~3 mins | ~2 seconds | ~8 seconds |
| 16 | ~12 mins | ~7 seconds | ~30 seconds |

### 7.2.3 Heuristic Runtime Analysis

The GALD heuristic was limited to a runtime of approximately one hour (epoch length) for all experiments. The ability to change P-states via DVFS gives the GALD a powerful advantage over the FDLD heuristic (which does not take advantage of different P-states in compute nodes) but it also means that GALD takes longer to execute. Table 4 shows the approximate execution time of each heuristic for various problem sizes. The FDLD heuristic does not show better cost reduction than GALD but it has the advantage of reaching a solution more quickly. However, NSLD and NILD took a few seconds to execute. This is beneficial in cases where the workload manager must make the allocation decisions quickly. The NSLD heuristic reached the solution quicker than NILD because it uses a simplistic data center cost model. The runtime of FDLD, NSLD, and NILD can be further reduced by running simulations on powerful machines.

### 7.3 Task Arrival Rate Pattern Analysis

For this experiment, we considered a configuration with eight data centers executing both sinusoidal and flat workload arrival patterns as shown in Fig. 3. We conducted 10 simulation runs for each workload pattern and analyzed its impact. For each run, the arrival rate values were randomly sampled (with a normal distribution). We used the original arrival rate values, shown in Fig. 3 as mean and used 20% of the mean as standard deviation. We analyzed the cloud operating cost variation for each heuristic over a day by plotting the mean cost with the standard error bars as shown in Fig. 7. Recall that each task type is characterized by its arrival rate and the estimated time required to complete the task on each of the heterogeneous compute nodes. The heuristics distribute the workload to minimize total energy cost across all data centers with the constraint that the execution rates of all task types meet their arrival rates (2). Therefore, the workload assignment was altered with the change in incoming arrival rate pattern, which further affected the system cost. The results are shown in Fig. 7(a) and Fig. 7(b) indicate that the geo-distributed system responded differently for sinusoidal and flat arrival rate patterns. The percentage difference between sinusoidal and flat arrival rate patterns' results for 'PS, NM, NW aware' variants (yellow bars) of FDLD, NSLD, GALD, and NILD are 16%, 7%, 1%, and 10%, respectively.

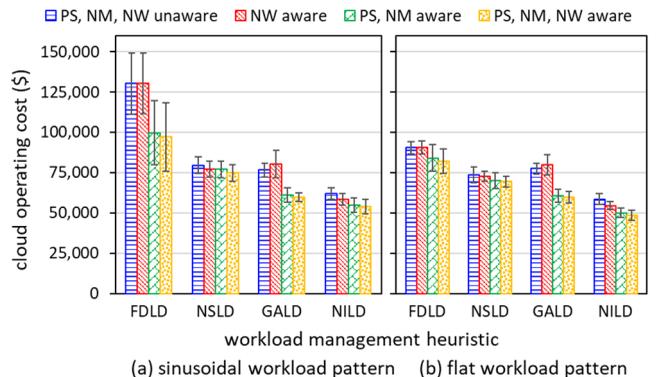

(a) sinusoidal workload pattern    (b) flat workload pattern

Fig. 7. Comparison of cloud operating costs among heuristics over a day for (a) sinusoidal and (b) flat workload arrival rate patterns for eight data centers

Overall cloud operating cost was higher for the sinusoidal workload arrival pattern because it produced higher peak demand costs and consumed the most renewable power available during the day as compared to the flat workload arrival pattern. The wide standard deviation for FDLD variants shows that the FDLD heuristic is sensitive to the varying arrival rates. For FDLD, slight variations in arrival rates change final allocations significantly that affect peak power costs. This causes the operating cost to fluctuate significantly. The NSLD and GALD heuristics are not as sensitive as FDLD, while NILD is the most consistent among all heuristics. We also notice that some simulation results for all three heuristics show that the NW aware heuristic variants perform worse than their NW unaware counterparts. This happens mostly with the GALD heuristic executing sinusoidal workload pattern. In such cases, the heuristic only considers the network costs while making allocation decisions and does not consider the peak shaving and net metering. This increases the peak demand costs and also does not allow heuristics to sell the excess energy back increasing the overall cloud operating cost.

## 7.4 Sensitivity Analysis

### 7.4.1 Task Data Size

We performed this experiment to analyze the impact of task data size on the cloud operating costs. Recall that, as per (6), two principal components of the network cost are the price per data traffic unit ($/GB), $N^{price}$, and the amount of data volume (GB) for the number of tasks migrated (outward). Here the data volume depends on the amount of data migrated per task. If a task migrates from one data center location to another, it transfers various types and amounts of data for each task type. We used a workload that was a mixture of offline analytics and artificial intelligence task types as shown in Table 2. This experiment considered all heuristics for a group of eight data centers executing both sinusoidal and flat workloads with different task data sizes. We considered the 'PS, NM, NW aware' variants of all workload management heuristics in this study.

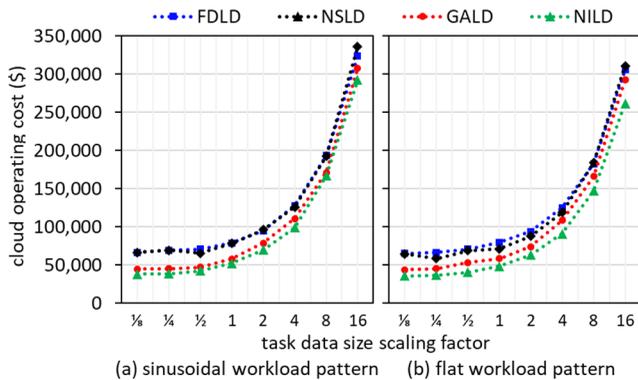

(a) sinusoidal workload pattern   (b) flat workload pattern

Fig. 8. Impact of task data size scaling on cloud operating costs among heuristics over a day for (a) sinusoidal and (b) flat workload arrival rate patterns for eight data centers

The results from Fig. 8 show that the cloud operating costs increased with the increase in task data size. Here, the network costs increased with an increase in the amount of data transferred. For all task data sizes, our proposed NILD framework performed the best overall. For the small values of task data sizes, both FDLD and NSLD performed the worst, while GALD performed better but not as good as NILD. For the large values of task data sizes, GALD's performance degraded (as compared to its performance for the small data sizes) but NSLD performed the worst.

### 7.4.2 NILD Delay Cost Factor (β)

As discussed in Section 5.5, NILD uses the *Best-Reply* algorithm to determine arrival rates. It uses the delay cost factor, $\beta$, while calculating the arrival rates (26). In this experiment, we study the impact of $\beta$ on the cloud operating cost. We considered configurations with four, eight, and sixteen data centers executing both sinusoidal and flat workload arrival patterns. We considered the 'PS, NM, NW unaware' variants of all workload management heuristics in this study.

Results from this experiment in Fig. 9 show that the cloud operating costs for both sinusoidal and flat workloads increased with the increase in $\beta$. As per (26), the larger values of $\beta$ make large adjustments in the arrival rate, and vice versa. For the configuration with sixteen data centers, when we increased $\beta$ beyond 0.2, the NILD failed to reach equilibrium. Whereas for the other configurations, NILD reached equilibrium quickly and CWM could not make fine adjustments in the arrival rate to further minimize the operating costs. When we decreased the value of $\beta$ very low (smaller than 0.1), NILD reached equilibrium slowly and could not make appropriate adjustments in the arrival rates. The best value of $\beta$ found from this experiment (0.1) was the one we used for all experiments.

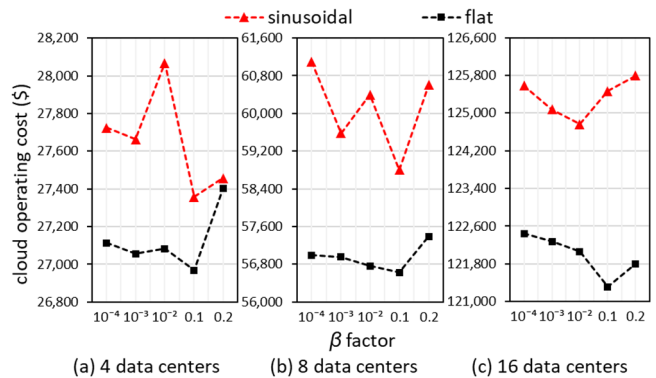

(a) 4 data centers   (b) 8 data centers   (c) 16 data centers

Fig. 9. Impact of the delay cost factor, β, on cloud operating costs among heuristics over a day for both sinusoidal and flat workload arrival rate patterns, for a configuration with (a) four, (b) eight, and (c) sixteen data centers

## 8 CONCLUSIONS

In this work, we studied the problem of workload distribution among geo-distributed cloud data centers to minimize cloud energy and network costs while ensuring all tasks complete without being dropped. We used a new cloud network model that directly considers the real-world data transfer prices and the amount of data migrated out of the cloud data centers. Our framework considers heterogeneity within the data centers and in task types used in the workload. It is aware of data center cooling power, time-of-use (TOU) electricity pricing, green renewable energy, net metering, peak demand pricing distribution, inter-data center network, and data center queueing delay. We formulated the cloud workload distribution problem as a non-cooperative game. We proposed a game-theoretic workload management technique for minimizing the cloud operating cost. However, to implement our approach in a real system, it needs to be used with some form of a workload prediction technique, e.g., [13], [14], to avoid delays with workload allocation. For our complex (NP-hard) problem and system environment, it should also be noted that the Nash equilibrium-based game-theoretic technique may not always achieve global optima solutions. However, the obtained solutions are still superior to those obtained with state-of-the-art frameworks, as demonstrated in our experiments. Apart from the data analytics and artificial intelligence workloads, the real-world cloud workloads also include web-service, search, mobile services, etc. Such time-critical workloads cannot be transferred without migration penalties, extra network processing, and scheduling costs. However, our framework can be extended for such time-critical workloads, e.g., by using a priority scheduling approach where time-critical workloads are given higher priority for local scheduling at (or near) the source of the request, to minimize latencies. We compared our new game-theoretic NILD technique with three state-of-the-art techniques (FDLD, GALD [36], and NSLD [35]). We analyzed their performance by comparing the cloud operating cost reduction, performing a scalability assessment, examining sensitivity to task data size, testing system behavior for different task arrival patterns, and comparing data center queueing delays. We demonstrated that NILD performed the best when including information about peak demand charges, net metering policies, network costs, and intra-data center queueing delay. The best performing NILD heuristic from this work achieved 43%, 16%, and 33% cost reductions on average than the FDLD and GALD from [36], and NSLD from [35], respectively. Additionally, the runtime of the NILD heuristic is much lower than the FDLD and GALD heuristics, and is suitable in case of time-critical workload scheduling.

## ACKNOWLEDGMENTS

The authors thank Dylan Machovec for his valuable comments on this work. This work is supported by the National Science Foundation (NSF) under grant CCF-1302693.

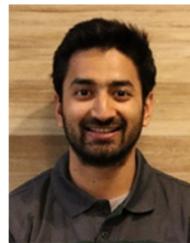

**Ninad Hogade** received his B.E. degree in Electronics Engineering from Vishwakarma Institute of Technology, India, and M.S. degree in Computer Engineering from Colorado State University, USA. He is currently a Ph.D. student in Computer Engineering at Colorado State University, USA. His research interests include energy aware scheduling of high performance computing systems and data centers.

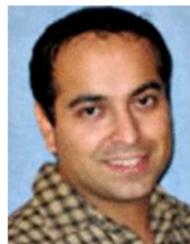

**Sudeep Pasricha** received his B.E. degree in Electronics and Communications from Delhi Institute of Technology, India; and his M.S. and Ph.D. degrees in Computer Science from University of California, Irvine. He is currently a Professor and Chair of Computer Engineering at Colorado State University, where he is also a Professor of Computer Science and Systems Engineering. He is a Senior Member of the IEEE and an ACM Distinguished Member. Homepage: http://www.engr.colostate.edu/sudeep.

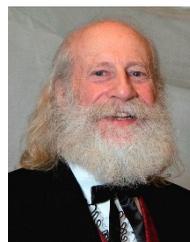

**Howard Jay (H.J.) Siegel** is a Professor Emeritus at Colorado State University. From 2001 to 2017, he was the Abell Endowed Chair Distinguished Professor of Electrical and Computer Engineering, and a Professor of Computer Science. He was a professor at Purdue from 1976 to 2001. He is an IEEE Fellow and an ACM Fellow. He received B.S. degrees from MIT, and the M.A., M.S.E., and Ph.D. degrees from Princeton.